\documentclass[sigconf,screen,authorversion,nonacm]{acmart}

\AtBeginDocument{%
  }

\copyrightyear{2024}
\acmYear{2024}
\setcopyright{rightsretained}
\acmConference[SPAA '24]{Proceedings of the 36th ACM Symposium on Parallelism in Algorithms and Architectures}{June 17--21, 2024}{Nantes, France}
\acmBooktitle{Proceedings of the 36th ACM Symposium on Parallelism in Algorithms and Architectures (SPAA '24), June 17--21, 2024, Nantes, France}\acmDOI{10.1145/3626183.3660257}
\acmISBN{979-8-4007-0416-1/24/06}

\usepackage{tikz}
\usetikzlibrary{external}
\usepackage{algorithmic}
\usepackage{xcolor}
\usepackage{mathtools}
\usepackage{cleveref}
\usepackage{numprint}
\npdecimalsign{.}
\npthousandsep{\,}
\usepackage[algo2e, ruled, vlined, linesnumbered]{algorithm2e}
\usepackage{booktabs}
\usepackage{multirow}
\usepackage{wasysym}

\newcommand{\myparagraph}[1]{\paragraph{#1}}

\newcommand{\cupeqq}{\mathrel{\cup}=}

\SetKw{KwBreak}{break}
\SetKwInput{KwInvariant}{invariant}
\SetKw{KwWhile}{while}
\SetKw{KwContinue}{continue}

\DeclareMathOperator*{\argmax}{arg\,max}
\DeclareMathOperator*{\argmin}{arg\,min}

\newcommand{\symbTimeout}{\clock}
\newcommand{\symbInfeasible}{\ensuremath{\times}}

\newcommand{\modified}[1]{{\color{red}#1}}
\renewcommand{\modified}[1]{#1}

\newcommand{\GmeanCutImprMtKaHyParOverContrib}{3.5}

\newif\ifappendix
\appendixtrue

\begin{document}

\title{Brief Announcement: Distributed Unconstrained Local Search for Multilevel Graph Partitioning}

\author{Peter Sanders}
\email{sanders@kit.edu}
\affiliation{%
  \institution{Karlsruhe Institute of Technology}
  \city{Karlsruhe}
  \country{Germany}
}

\author{Daniel Seemaier}
\email{daniel.seemaier@kit.edu}
\affiliation{%
  \institution{Karlsruhe Institute of Technology}
  \city{Karlsruhe}
  \country{Germany}}

\begin{abstract}
    Partitioning a graph into blocks of \emph{roughly equal} weight while cutting only few edges is a fundamental problem in computer science with numerous practical applications.
    While shared-memory parallel partitioners have recently matured to achieve the same quality as widely used sequential partitioners, there is still a pronounced quality gap between distributed partitioners and their sequential counterparts.
    In this work, we shrink this gap considerably by describing the engineering of an unconstrained local search algorithm suitable for distributed partitioners.
    We integrate the proposed algorithm in a distributed multilevel partitioner.
    Our extensive experiments show that the resulting algorithm scales to thousands of PEs while computing cuts that are, on average, only 3.5\% larger than those of a state-of-the-art high-quality shared-memory partitioner.
    Compared to previous distributed partitioners, we obtain on average 6.8\% smaller cuts than the best-performing competitor while being more than 9 times faster.
\end{abstract}

\begin{CCSXML}
<ccs2012>
<concept>
<concept_id>10003752.10003809.10010172</concept_id>
<concept_desc>Theory of computation~Distributed algorithms</concept_desc>
<concept_significance>500</concept_significance>
</concept>
<concept>
<concept_id>10003752.10003809.10010170</concept_id>
<concept_desc>Theory of computation~Parallel algorithms</concept_desc>
<concept_significance>500</concept_significance>
</concept>
</ccs2012>
\end{CCSXML}

\ccsdesc[500]{Theory of computation~Distributed algorithms}
\ccsdesc[500]{Theory of computation~Parallel algorithms}
\keywords{graph partitioning, distributed algorithms, parallel algorithms}

\maketitle

\section{Introduction}
\label{s:introduction}
Graphs are a central concept of computer science used whenever we need to model relations between objects.
Consequently, handling \emph{large} graphs is very important for parallel processing. 
This often requires to \emph{partition} these graphs into blocks of roughly equal weight while cutting only few edges between the blocks (\emph{balanced} graph partitioning).

More precisely, consider a graph $G = (V = 1..n, E, c, \omega)$ with positive vertex weights $c$ and edge weights $\omega$.
We are looking for a partition $\Pi$ of $V$ into $k$ non-overlapping blocks $V_1, \dots, V_k$.
The \emph{balance constraint} demands that for all blocks $c(V_i) \le L_{\textrm{max}} \coloneqq (1 + \varepsilon) \frac{c(V)}{k}$ for some imbalance parameter $\varepsilon$.
The objective is to minimize the \emph{cut}, i.e., the total weight of all inter-block edges.

There has been a huge amount of research on graph partitioning, so we refer the reader to overview papers \cite{MoreRecentAdvances} for most of the general material.
\emph{Multilevel graph partitioners} (e.g., \cite{ParMETIS, ParHIP, Mt-KaHyPar-D}) achieve high-quality partitions for a wide range of input graphs with a good trade-off between quality and partitioning cost.
First, iteratively \emph{coarsen} the graph $G$.
The resulting small graph $G'$ is a good representation of the input and an \emph{initial partition} of $G'$ already induces a good partition of $G$.
This is further improved by \emph{uncoarsening} the graph and improving the partition on each level through refinement algorithms.
Arguably the most successful refinement algorithm is the \emph{Fiduccia-Mattheyses} local search algorithm~\cite{FM} (FM).

However, parallelizing the refinement phase has proved challenging over several decades.
Shared-memory partitioners have recently matured to achieve high quality and reasonable scalability~\cite{Mt-KaHyPar-D} through an efficient parallelization of the FM algorithm, requiring fine-grained synchronization and communication, which is unsuitable for distributed memory.
On the other hand, distributed partitioners~\cite{ParHIP, ParMETIS} rely on simpler local search algorithms such as label propagation, inducing a considerable quality penalty.
Label propagation visits each vertex in parallel and moves the vertex to the block that minimizes the induced cut while preserving the balance constraint.
This process is usually repeated for multiple rounds and terminated once no more vertices are moved during a round (i.e., the partition is in a local optimum) or a maximum number of rounds has been exceeded.

Recently, Gilbert~et~al.~\cite{Jet} introduced the refinement algorithm \emph{Jet}, which they show to achieve similar partition quality to FM refinement while being suitable for GPU parallelism. 
The Jet algorithm first builds a set of \emph{move candidates} $M$: for each vertex $v$ in block $V_i$, let $\text{conn}(v, V')$ be the summed edge weight between $v$ and vertices in block $V'$.
Let $V_j = \argmax_{V'} \text{conn}(v, V')$ be the block with the strongest connection to $v$.
Then, $v$ is included in $M$ if $\text{conn}(v, V_j) - \text{conn}(v, V_i) \ge - \lfloor \tau \cdot \text{conn}(v, V_i) \rfloor$, where $\tau$ is a constant \emph{temperature} parameter.
For $\tau = 1$, all vertices are included, whereas $\tau = 0$ only considers vertices with positive gain.
Note that in contrast to \emph{standard} label propagation, $M$ might also include moves that violate the balance constraint.
The moves in $M$ are then sorted by their gain value before positive gain moves are applied, locking them in the next round to avoid oscillation. 
If violated, an independent \emph{rebalancing} algorithm fixes the balance constraint afterward.

Based on the idea of balance-violating moves with subsequent rebalancing, Maas~et~al.~\cite{Mt-KaHyPar-UFM} developed an \emph{unconstrained FM} adaptation as part of the shared-memory parallel partitioner \textsf{Mt-KaHyPar}.
Their experimental evaluation shows that unconstrained FM achieves considerable improvements compared to regular FM (and also Jet) on real-world graphs with a power-law degree distribution. 

\ifappendix
\else
\vspace{-0.5em}
\fi
\myparagraph{Contributions.}
We achieve similar quality and much better scalability by porting Jet to a distributed-memory setting. 
Move generation is relatively easy to adapt but further improved by working with multiple temperatures. 
We develop an alternative highly scalable probabilistic algorithm for rebalancing, that can perform a large number of favorable balancing moves in parallel.
We finish up using the slower but more controlled algorithm from dKaMinPar~\cite{dKaMinPar}.
It remains to be seen whether this approach is more suitable if $k$ is large.
Extensive experiments on real-world graphs and huge randomly generated networks indicate scalability to (at least) 8\,192 cores and one trillion (directed) edges.
Compared to state-of-the-art distributed partitioners we achieve considerably improved quality when partitioning into $k \in \{2, 4, \dots, 128\}$ blocks.
\ifappendix
\else
\vspace{-0.5em}
\fi
\section{Distributed Jet Refinement}\label{s:jet}

We consider a graph $G$ distributed across $P$ processing elements (PEs) $1..P$, each containing a subgraph of consecutive vertices of $G$ distributed s.t. each PE has roughly the same \emph{number of edges}.
Undirected edges $\{u, v\}$ are represented by two directed edges $(u, v)$ and $(v, u)$, which are stored on the PEs owning the respective tail vertices.
Vertices adjacent to vertices owned by other PEs are called \emph{interface vertices} and replicated as \emph{ghost vertices} (i.e., without outgoing edges).
The partition $\Pi = \{V_1, \dots, V_k\}$ is distributed s.t. each PE stores the block IDs for its vertices, including its ghost vertices.

\ifappendix
\else
\vspace{-0.5em}
\fi
\myparagraph{Jet Refinement.}
Implementing Jet~\cite{Jet} in this model requires only minor modifications.
We restate the algorithm here in order to be self-contained.
First, recall that each PE builds a set of local move candidates $M$ 
by visiting its owned 
vertices, including a vertex $v$ of block $V_i$ in $M$ if $g(v) \coloneqq \max_{j \neq i} \text{conn}(v, V_j) - \text{conn}(v, V_i) \ge - \lfloor \tau \cdot \text{conn}(v, V_i) \rfloor$.
The temperature $\tau$ controls the extent of move candidates with negative gain (i.e., that would increase the partition cut) included in $M$.
Note that $\text{conn}(v, \cdot)$ can be computed without communication since it only depends on $v$'s neighborhood.
Next, each interface vertex $v \in M$ sends its corresponding $g(v)$ value to its ghost replicates.
With this information, PEs independently re-evaluate their move candidates, removing all vertices $v$ that would increase the partition cut assuming that any neighbor $u$ with $g(u) > g(v)$ is moved before $v$.
The remaining vertices are moved to their target blocks, locked for the next round, and the block assignment of ghost replicates is updated accordingly.

\ifappendix
\else
\vspace{-0.5em}
\fi
\myparagraph{Rebalancing.}
Rebalancing the partition afterwards while retaining most of its improved quality is crucial to the overall performance of the algorithm.
Ref.~\cite{dKaMinPar} introduces a distributed rebalancing algorithm that can only move a few vertices per overloaded block in every centrally coordinated epoch.
This induces a sequential bottleneck for partitions with highly overloaded blocks.
\ifappendix
To make rebalancing scalable, we extend the algorithm by the following highly parallel approach (summarized in Algorithm~\ref{alg:rebalance}, \Cref{apx:pseudocode}) whenever a single round reduces the total partition overload by less than $10\%$.
\else
To make rebalancing scalable, we extend the algorithm by a round of the following highly parallel approach whenever a single round reduces the total partition overload by less than $10\%$.
\fi

Following Ref.~\cite{dKaMinPar}, we consider vertices $v \in V_o$ of overloaded blocks $V_o \in \Pi$ and look at the maximum reduction in edge cut $g_v$ when moving $v$ to any non-overloaded block.
Then, each PE sorts its vertices into exponentially spaces buckets $b^j_o$ based on their \emph{maximum relative gain}, defined as $r_v \coloneqq g_v \cdot c(v)$ if $g_v > 0$ and $r_v \coloneqq g_v / c(v)$ otherwise.
A vertex $v$ with $r_v \ge 0$ is assigned to bucket $j = 0$, whereas the bucket of negative gain vertices is calculated as $j = 1 + \lceil \log_{\alpha} (1 - r_v) \rceil$.
During early experimentation, we observed that the choice of $\alpha$ considerably impacts the resulting partition quality after rebalancing.
Thus, we use $\alpha = 1.1$ to obtain our results.
After constructing the PE-local buckets, we compute global bucket weights $c(B^i_o) \coloneqq \sum_p c(b^i_o @ p)$ using an all-reduce operation, where $b^i_o @ p$ refers to the value of variable $b^i_o$ on PE $p$.
The \emph{cut-off buckets} $\hat{B}_o \coloneqq \min \left\{ j \mid \sum_{i < j} c(B^i_o) \ge c(V_o) - L_{\max} \right\}$ contain the lowest-rated vertices that must be moved to remove all excess weight from overloaded blocks.
Since moving all vertices in buckets up to the cut-off buckets could introduce overloaded target blocks, we use a probabilistic approach that maintains balance (of non-overloaded blocks) in expectation.\footnote{Note that a more straightforward adaptation of the rebalancing algorithm proposed by Ref.~\cite{Jet} to distributed memory could work as follows. As proposed in Ref.~\cite{Jet}, only consider moves to target blocks with weights below the \emph{dead zone} $L_{\text{max}} - \delta \cdot (L_\text{max} - c(V) / k)$, where $\delta \in [0, 1]$ is a tuning parameter. Then, for each overloaded block $V_o$, always apply all moves in buckets below the cut-off bucket $\hat{B}_o$. 
Additionally, let PE $p$ move the highest rated vertices of the cut-off bucket $\hat{B}_o$ with total weight proportional to the total weight of block $V_o$ on PE $p$.
While this process could introduce new overloaded blocks, it terminates in $\le k$ iterations as at least one block per round is moved into the dead zone. 
It is open to see which approach gives better performance, especially for large $k$.}
To this end, let $W_u$ be the total vertex weight of all move candidates in buckets up to the cut-off bucket
with target block $V_u$.
Then, move vertices to $V_u$ with probability $p_u \coloneqq (L_{\max} - c(V_u)) / W_u$
(if $p_u \ge 1$, we always perform the move).

\ifappendix
\else
\vspace{-0.5em}
\fi
\myparagraph{Integration.}
We perform $t = 4$ rounds of the proposed algorithm.
Since we have observed that Jet is very sensitive to the temperature used in the construction of $M$, we perform multiple rounds of the algorithm to achieve more robust performance.
During the $i$-th round, we set the temperature to $\tau_i = \tau_0 + \frac{i}{t} (\tau_1 - \tau_0)$, where $\tau_0 = 0.75$ and $\tau_1 = 0.25$.
Following Ref.~\cite{Jet}, a single round consists of repeating the described Jet refinement and rebalancing steps until 12 consecutive repetitions did not improve the partition.

\ifappendix
\else
\vspace{-0.5em}
\fi
\section{Experiments}\label{s:experiments}

\begin{figure*}[t]
    \centering
    \input{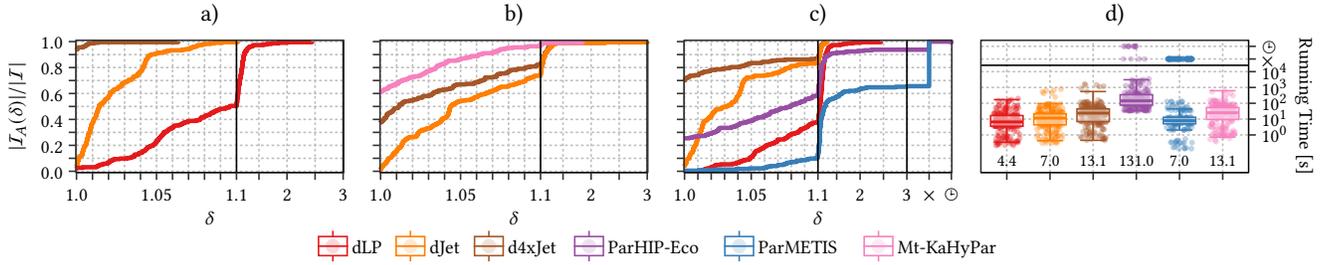}
    % Created by tikzDevice version 0.12.4 on 2024-04-23 13:03:18
% !TEX encoding = UTF-8 Unicode
\begin{tikzpicture}[x=1pt,y=1pt]
\definecolor{fillColor}{RGB}{255,255,255}
\begin{scope}
\definecolor{fillColor}{RGB}{255,255,255}

\path[fill=fillColor] (119.57,245.72) rectangle (134.03,260.17);
\end{scope}
\begin{scope}
\definecolor{drawColor}{RGB}{228,26,28}
\definecolor{fillColor}{RGB}{228,26,28}

\path[draw=drawColor,draw opacity=0.33,line width= 0.4pt,line join=round,line cap=round,fill=fillColor,fill opacity=0.33] (126.80,252.94) circle (  2.50);
\end{scope}
\begin{scope}
\definecolor{drawColor}{RGB}{228,26,28}

\path[draw=drawColor,line width= 0.6pt,line join=round] (121.02,252.94) -- (132.58,252.94);
\end{scope}
\begin{scope}
\definecolor{drawColor}{RGB}{228,26,28}

\path[draw=drawColor,line width= 0.6pt] (126.80,247.16) --
	(126.80,249.33);

\path[draw=drawColor,line width= 0.6pt] (126.80,256.56) --
	(126.80,258.73);
\definecolor{fillColor}{RGB}{255,255,255}

\path[draw=drawColor,line width= 0.6pt,fill=fillColor,fill opacity=0.50] (121.38,249.33) rectangle (132.22,256.56);

\path[draw=drawColor,line width= 0.6pt] (121.38,252.94) --
	(132.22,252.94);
\end{scope}
\begin{scope}
\definecolor{fillColor}{RGB}{255,255,255}

\path[fill=fillColor] (148.91,245.72) rectangle (163.36,260.17);
\end{scope}
\begin{scope}
\definecolor{drawColor}{RGB}{255,127,0}
\definecolor{fillColor}{RGB}{255,127,0}

\path[draw=drawColor,draw opacity=0.33,line width= 0.4pt,line join=round,line cap=round,fill=fillColor,fill opacity=0.33] (156.14,252.94) circle (  2.50);
\end{scope}
\begin{scope}
\definecolor{drawColor}{RGB}{255,127,0}

\path[draw=drawColor,line width= 0.6pt,line join=round] (150.36,252.94) -- (161.92,252.94);
\end{scope}
\begin{scope}
\definecolor{drawColor}{RGB}{255,127,0}

\path[draw=drawColor,line width= 0.6pt] (156.14,247.16) --
	(156.14,249.33);

\path[draw=drawColor,line width= 0.6pt] (156.14,256.56) --
	(156.14,258.73);
\definecolor{fillColor}{RGB}{255,255,255}

\path[draw=drawColor,line width= 0.6pt,fill=fillColor,fill opacity=0.50] (150.72,249.33) rectangle (161.56,256.56);

\path[draw=drawColor,line width= 0.6pt] (150.72,252.94) --
	(161.56,252.94);
\end{scope}
\begin{scope}
\definecolor{fillColor}{RGB}{255,255,255}

\path[fill=fillColor] (178.58,245.72) rectangle (193.04,260.17);
\end{scope}
\begin{scope}
\definecolor{drawColor}{RGB}{166,86,40}
\definecolor{fillColor}{RGB}{166,86,40}

\path[draw=drawColor,draw opacity=0.33,line width= 0.4pt,line join=round,line cap=round,fill=fillColor,fill opacity=0.33] (185.81,252.94) circle (  2.50);
\end{scope}
\begin{scope}
\definecolor{drawColor}{RGB}{166,86,40}

\path[draw=drawColor,line width= 0.6pt,line join=round] (180.03,252.94) -- (191.59,252.94);
\end{scope}
\begin{scope}
\definecolor{drawColor}{RGB}{166,86,40}

\path[draw=drawColor,line width= 0.6pt] (185.81,247.16) --
	(185.81,249.33);

\path[draw=drawColor,line width= 0.6pt] (185.81,256.56) --
	(185.81,258.73);
\definecolor{fillColor}{RGB}{255,255,255}

\path[draw=drawColor,line width= 0.6pt,fill=fillColor,fill opacity=0.50] (180.39,249.33) rectangle (191.23,256.56);

\path[draw=drawColor,line width= 0.6pt] (180.39,252.94) --
	(191.23,252.94);
\end{scope}
\begin{scope}
\definecolor{fillColor}{RGB}{255,255,255}

\path[fill=fillColor] (216.48,245.72) rectangle (230.93,260.17);
\end{scope}
\begin{scope}
\definecolor{drawColor}{RGB}{152,78,163}
\definecolor{fillColor}{RGB}{152,78,163}

\path[draw=drawColor,draw opacity=0.33,line width= 0.4pt,line join=round,line cap=round,fill=fillColor,fill opacity=0.33] (223.70,252.94) circle (  2.50);
\end{scope}
\begin{scope}
\definecolor{drawColor}{RGB}{152,78,163}

\path[draw=drawColor,line width= 0.6pt,line join=round] (217.92,252.94) -- (229.48,252.94);
\end{scope}
\begin{scope}
\definecolor{drawColor}{RGB}{152,78,163}

\path[draw=drawColor,line width= 0.6pt] (223.70,247.16) --
	(223.70,249.33);

\path[draw=drawColor,line width= 0.6pt] (223.70,256.56) --
	(223.70,258.73);
\definecolor{fillColor}{RGB}{255,255,255}

\path[draw=drawColor,line width= 0.6pt,fill=fillColor,fill opacity=0.50] (218.28,249.33) rectangle (229.12,256.56);

\path[draw=drawColor,line width= 0.6pt] (218.28,252.94) --
	(229.12,252.94);
\end{scope}
\begin{scope}
\definecolor{fillColor}{RGB}{255,255,255}

\path[fill=fillColor] (273.28,245.72) rectangle (287.73,260.17);
\end{scope}
\begin{scope}
\definecolor{drawColor}{RGB}{55,126,184}
\definecolor{fillColor}{RGB}{55,126,184}

\path[draw=drawColor,draw opacity=0.33,line width= 0.4pt,line join=round,line cap=round,fill=fillColor,fill opacity=0.33] (280.50,252.94) circle (  2.50);
\end{scope}
\begin{scope}
\definecolor{drawColor}{RGB}{55,126,184}

\path[draw=drawColor,line width= 0.6pt,line join=round] (274.72,252.94) -- (286.28,252.94);
\end{scope}
\begin{scope}
\definecolor{drawColor}{RGB}{55,126,184}

\path[draw=drawColor,line width= 0.6pt] (280.50,247.16) --
	(280.50,249.33);

\path[draw=drawColor,line width= 0.6pt] (280.50,256.56) --
	(280.50,258.73);
\definecolor{fillColor}{RGB}{255,255,255}

\path[draw=drawColor,line width= 0.6pt,fill=fillColor,fill opacity=0.50] (275.08,249.33) rectangle (285.92,256.56);

\path[draw=drawColor,line width= 0.6pt] (275.08,252.94) --
	(285.92,252.94);
\end{scope}
\begin{scope}
\definecolor{fillColor}{RGB}{255,255,255}

\path[fill=fillColor] (325.96,245.72) rectangle (340.42,260.17);
\end{scope}
\begin{scope}
\definecolor{drawColor}{RGB}{247,129,191}
\definecolor{fillColor}{RGB}{247,129,191}

\path[draw=drawColor,draw opacity=0.33,line width= 0.4pt,line join=round,line cap=round,fill=fillColor,fill opacity=0.33] (333.19,252.94) circle (  2.50);
\end{scope}
\begin{scope}
\definecolor{drawColor}{RGB}{247,129,191}

\path[draw=drawColor,line width= 0.6pt,line join=round] (327.41,252.94) -- (338.97,252.94);
\end{scope}
\begin{scope}
\definecolor{drawColor}{RGB}{247,129,191}

\path[draw=drawColor,line width= 0.6pt] (333.19,247.16) --
	(333.19,249.33);

\path[draw=drawColor,line width= 0.6pt] (333.19,256.56) --
	(333.19,258.73);
\definecolor{fillColor}{RGB}{255,255,255}

\path[draw=drawColor,line width= 0.6pt,fill=fillColor,fill opacity=0.50] (327.77,249.33) rectangle (338.61,256.56);

\path[draw=drawColor,line width= 0.6pt] (327.77,252.94) --
	(338.61,252.94);
\end{scope}
\begin{scope}
\definecolor{drawColor}{RGB}{0,0,0}

\node[text=drawColor,anchor=base west,inner sep=0pt, outer sep=0pt, scale=  0.80] at (134.03,250.19) {dLP};
\end{scope}
\begin{scope}
\definecolor{drawColor}{RGB}{0,0,0}

\node[text=drawColor,anchor=base west,inner sep=0pt, outer sep=0pt, scale=  0.80] at (163.36,250.19) {dJet};
\end{scope}
\begin{scope}
\definecolor{drawColor}{RGB}{0,0,0}

\node[text=drawColor,anchor=base west,inner sep=0pt, outer sep=0pt, scale=  0.80] at (193.04,250.19) {d4xJet};
\end{scope}
\begin{scope}
\definecolor{drawColor}{RGB}{0,0,0}

\node[text=drawColor,anchor=base west,inner sep=0pt, outer sep=0pt, scale=  0.80] at (230.93,250.19) {ParHIP-Eco};
\end{scope}
\begin{scope}
\definecolor{drawColor}{RGB}{0,0,0}

\node[text=drawColor,anchor=base west,inner sep=0pt, outer sep=0pt, scale=  0.80] at (287.73,250.19) {ParMETIS};
\end{scope}
\begin{scope}
\definecolor{drawColor}{RGB}{0,0,0}

\node[text=drawColor,anchor=base west,inner sep=0pt, outer sep=0pt, scale=  0.80] at (340.42,250.19) {Mt-KaHyPar};
\end{scope}
\end{tikzpicture}
    \vspace{-1em}
    \caption{Solution quality comparison using performance profiles. %~\cite{DBLP:journals/mp/DolanM02}.
    Let $\mathcal{A}$ be the set of algorithms, $\mathcal{I}$ the set of instances, and $q_A(I)$ the edge cut of algorithm $A \in \mathcal{A}$ on instance $I \in \mathcal{I}$. 
    For each algorithm $A$, we plot the fraction of instances $\frac{\lvert \mathcal{I}_A(\delta) \rvert}{\lvert \mathcal{I} \rvert}$ (y-axis) where $\mathcal{I}_A(\delta) \coloneqq \{ I \in \mathcal{I} \mid q_A \le \delta \cdot \min_{A' \in \mathcal{A}} q_{A'}(I) \}$ and $\delta$ on the x-axis. 
    (a) Solution quality of plain \textsf{dKaMinPar} (\textsf{dLP}) and with 1 resp. 4 rounds of Jet refinement (\textsf{dJet} resp. \textsf{d4xJet}).
    (b) Comparison against \textsf{Mt-KaHyPar}~\cite{Mt-KaHyPar-UFM} (unconstrained FM refinement).
    (c) Comparison against distributed partitioners \textsf{ParHIP}~\cite{ParHIP} and \textsf{ParMETIS}~\cite{ParMETIS}.
    (d) Per-instance running times with gmeans across all instances that were solved by all algorithms or exceeded the time limit (1\,h).}
    \label{fig:quality}
    \vspace{-1.5em}
\end{figure*}

We have integrated the proposed refinement algorithm into the distributed multilevel partitioner \textsf{dKaMinPar}~\cite{dKaMinPar}, compiled all codes using g++-12.1 with flags \texttt{-O3 -march=native} and use OpenMPI 4.0 as parallelization library.
\ifappendix
We use the benchmark set of Ref.~\cite{dKaMinPar} (\Cref{apx:benchmark}) to evaluate the solution quality of the proposed refinement algorithm,
\else
We use the benchmark set of Ref.~\cite{dKaMinPar} (32 graphs) to evaluate the solution quality of the proposed refinement algorithm,
\fi
partitioning each graph using $k \in \{2, 4, \dots, 128\}$ and $\varepsilon = 3\%$ on a single machine equipped with a 64-core AMD EPYC 7702P and 1\,TB of main memory.

\ifappendix
Looking at Fig.~\ref{fig:quality}a, we observe that our strongest configuration with 4 rounds of Jet (denoted \textsf{d4xJet}) improves the partition cut by at least \modified{10\%} on roughly \modified{50\%} of all benchmark instances compared to plain \textsf{dKaMinPar}, which only uses label propagation for refinement.
This is considered a lot in the context of multilevel graph partitioning, as most multilevel partitioners achieve average edge cuts within a few percentage points of each other.
\else
Looking at Fig.~\ref{fig:quality}a, we observe that \textsf{d4xJet} improves the partition cut by at least \modified{10\%} on roughly \modified{50\%} of all instances compared to plain \textsf{dKaMinPar}, which only uses label propagation for refinement.
This is considered a lot in the context of multilevel graph partitioning, as most multilevel partitioners achieve average edge cuts within a few percentage points of each other.
\fi

Compared to the state-of-the-art shared-memory partitioner \textsf{Mt-KaHyPar}~\cite{Mt-KaHyPar-UFM}, we find that \textsf{d4xJet} computes partitions with at most \modified{1\%} larger cuts on \modified{50\%} of all instances, shrinking the quality gap between distributed and shared-memory partitioners considerably (Fig.~\ref{fig:quality}b).
On average, the cuts found by Mt-KaHyPar are \modified{\GmeanCutImprMtKaHyParOverContrib}\% smaller than those of \textsf{d4xJet}, with roughly equal running times (Fig.~\ref{fig:quality}d).% when run on a single compute node. 

Fig.~\ref{fig:quality}c compares against competing distributed partitioners.
\textsf{d4xJet} finds at least $\approx 8\%$ lower cuts on roughly 50\% of all instances than \textsf{ParHIP-Eco}, which obtained the best cuts in Ref.~\cite{dKaMinPar}. %, we observe that \textsf{d4xJet} finds at least \modified{8\%} lower cuts on roughly \modified{50\%} of all instances, with \modified{\GmeanCutImprContribOverParHIPEco\%} lower cuts on average.
More, \textsf{ParHIP-Eco} is considerably slower than \textsf{d4xJet} due to its expensive evolutionary initial partitioning algorithm.%since it spends a lot of time computing high-quality initial partitions using an evolutionary algorithm.

\begin{figure}[t]
    \centering
    \input{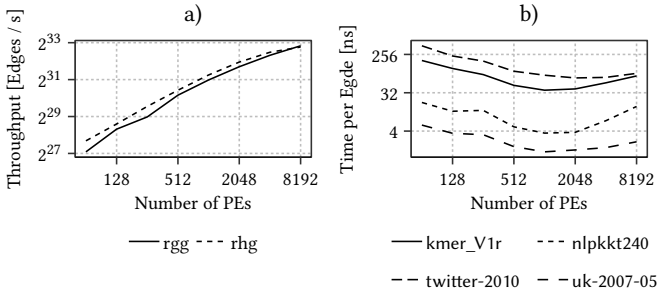}
    \vspace{-2.5em}
    \caption{Results for d4xJet. (a) Weak scaling on random geometric (rgg) and hyperbolic (rhg, power-law exponent $3.0$) graphs with $2^{21}$ vertices and $2^{27}$ (directed) edges per core.
    (b) Strong scaling on real-world graphs with low (\textsf{kmer\_V1r}, \textsf{nlpkkt240}) and high (\textsf{twitter-2010}, \textsf{uk-2007-05}) max. degree.}
    \label{fig:scaling}
    \ifappendix
    \vspace{-1.5em}
    \else
    \vspace{-1.5em}
    \fi
\end{figure}

To evaluate the scalability of \textsf{d4xJet}, we perform additional weak- and strong-scaling experiments on $\{1, 2, \dots, 128\}$ nodes of the HoreKa high-performance cluster, setting $k = 16$ and $\varepsilon = 3\%$.
Each node is equipped with two 38-core Intel Xeon Platinum 8368 processors\footnote{We only use 64 out of the available 76 cores because some of the graph generators require the number of cores to be a power of two.}.
The nodes are connected by an InfiniBand 4X HDR 200 Gbit/s network with approx. 1\,$\mu$s latency.
Results are shown in Fig.~\ref{fig:scaling}, where we observe weak scalability up to (at least) 8\,192 cores on randomly generated graphs.
On medium-sized real-world graphs, we generally observe scalability up to 1\,024--2\,048 cores.
This is roughly in line with competing distributed partitioners (see, e.g., \cite{dKaMinPar}), which we discuss briefly.
ParMETIS is unable to partition the graphs with high maximum degree due to its matching-based coarsening strategy, does not scale on \textsf{kmer\_V1r}, but scales up to 2\,048 cores on \textsf{nlpkkt240} with speedup \modified{8.7} (over \modified{64} cores; d4xJet scales to 1\,024 cores with speedup \modified{5.3}).
ParHIP(-Fast) fails to partition the \textsf{twitter-2010} and \textsf{kmer\_V1r} graphs regardless of the number of cores used and achieves a speedup of \modified{2.1} on \modified{2\,048} cores for \textsf{uk-2007-05} (d4xJet achieves \modified{3.9}), and \modified{4.5} on \modified{512} cores on \textsf{nlpkkt240}.
\ifappendix
Interestingly, we note that the edge cuts improve slightly when increasing the number of cores (see Appendix~\ref{apx:strongscaling_quality}).
\else
Interestingly, we note that edge cuts improve slightly when increasing the number of cores.%~\cite{FullPaperTODO}.
\fi

\ifappendix
\else
\vspace{-0.5em}
\fi
\section{Conclusion}\label{s:conclusion}

We have engineered a distributed implementation of the Jet algorithm~\cite{Jet}, shrinking the quality gap between distributed and shared-memory partitioners considerably while scaling to thousands of cores.
Further improvements might be possible by using more sophisticated local search algorithms. 
For instance, Pt-Scotch~\cite{Pt-Scotch} uses the FM algorithm on the \emph{band graph} induced by the partition (local copies of the subgraph surrounding the cut), but this imposes a sequential bottleneck on graphs with large cuts.
Another direction for future work could be a Jet-based partitioner for distributed GPUs.

\ifappendix
\else
\vspace{-0.5em}
\fi
\begin{acks}
    This project has received funding from the European Research Council (ERC) under the European Union’s Horizon 2020 research and innovation programme (grant agreement No. 882500).
    This work was performed on the HoreKa supercomputer funded by the Ministry of Science, Research and the Arts Baden-Württemberg and by the Federal Ministry of Education and Research. 

    \begin{center}\includegraphics[height=0.7cm]{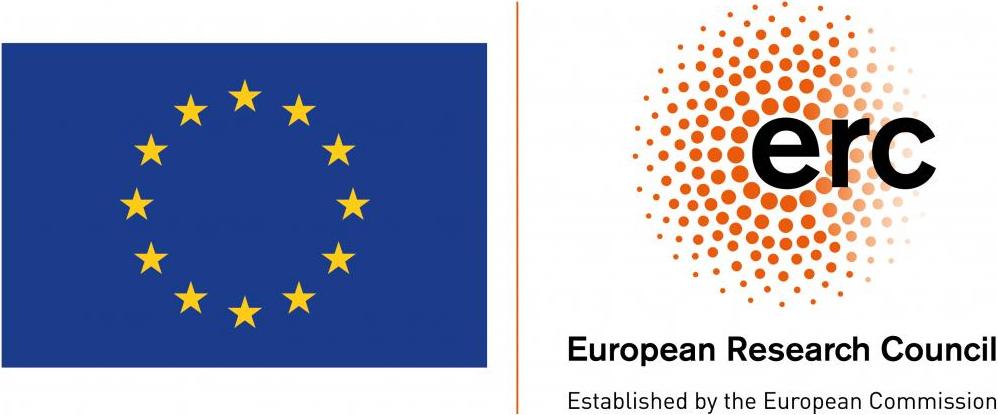}\end{center}
\end{acks}

\ifappendix
\else
\vspace{-2em}
\fi
\bibliographystyle{ACM-Reference-Format}
\bibliography{references}

\ifappendix
\appendix
\newpage

\section{Refinement Algorithm: Pseudocode}
\label{apx:pseudocode}

\begin{algorithm2e}[h]
    \KwIn{$G = (V, E, c, \omega)$, partition $\Pi$}
    \small{\tcp{1. Sort vertices into exponentially spaced buckets}}
    \ForAll{overloaded blocks $V_o \in \Pi$}{
        $b^i_o \coloneqq \emptyset$ for all $i$ \\
        \ForAll{$v \in V_o$}{
            $r_v \coloneqq \max_{\{V_u \in \Pi \mid c(V_u) + c(v) \le L_{\max}\}} \FuncSty{RelGain}(v, V_u)$ \\
            $j \coloneqq \left\{ 
            \begin{array}{lc} 
              0, & r_v \ge 0 \\ 
              1 + \lceil \log_{\alpha}(1 - r_v) \rceil , & r_v < 0 
            \end{array} \right.$ \\
            $b^j_o \cupeqq v$ \\
        }
    }
    \small{\tcp{2. Compute cut-offs, total weight to each block}}
    \ForAll{overloaded block $V_o \in \Pi$}{
        $c(B^i_o) \coloneqq \sum_p c(b^i_o @ p)$ for $i$ \\
        $\hat{B}_o \coloneqq \min \left\{ j \mid \sum_{i < j} c(B^i_o) \ge c(V_o) - L_{\max} \right\}$ \\
    }
    \ForAll{underloaded block $V_u \in \Pi$}{
        $W_u \coloneqq \sum_{V_o, p} c(\{v \in b^i_o @ p \mid i < \hat{B}_o \text{, } v \text{ moves to $V_u$}\})$ \\
        $p_u \coloneqq (L_{\max} - c(V_u)) / W_u$ \\
    }
    \small{\tcp{3. Move vertices}}
    \ForAll{overloaded block $V_o \in \Pi$, $j < \hat{B}_o$, $v \in b^j_o$}{
        \small{$V_u \coloneqq \argmin_{\{V_u \in \Pi \mid c(V_u) + c(v) \le L_{\max}\}} \FuncSty{RelGain}(v, V_u)$} \\
        Move $v$ from $V_o$ to $V_u$ with prob. $p_u$ \\
    }
    \Return{$\Pi$} \\
    \caption{$\FuncSty{Rebalance}(G, \Pi)$}
    \label{alg:rebalance}
\end{algorithm2e}

\section{Strong Scaling: Quality results}
\label{apx:strongscaling_quality}

\begin{table}[h]
    \small
    \centering
    \caption{Edge cuts obtained by \textsf{d4xJet} on real-world graphs using varying numbers of cores.}
    \begin{tabular}{l|rrrr}
        $P$ & \textsf{kmer\_V1r} & \textsf{nlpkkt240} & \textsf{twitter-2010} & \textsf{uk-2007-05} \\
        \midrule
        64 & 9\,366 & 4\,812 & 473\,004 & 3\,716 \\
        256 & 9\,240 & 4\,824 & 473\,277 & 3\,602 \\
        1\,024 & 9\,269 & 4\,811 & 472\,806 & 3\,575 \\
        4\,096 & \textbf{9\,055} & 4\,831 & \textbf{471\,340} & 3\,459 \\
        8\,192 & 9\,140 & \textbf{4\,738} & 471\,508 & \textbf{3\,424} \\
        \midrule
        \multicolumn{1}{c}{} & \multicolumn{4}{c}{cut $\times$ 1\,000}
    \end{tabular}
\end{table}

\newpage

\section{Benchmark Instances}
\label{apx:benchmark}

\begin{table}[h]
    \small
    \centering
    \caption{Basic properties of the benchmark set. Instances are roughly classified as \emph{low degree} and \emph{high degree} graphs based on their maximum degree $\Delta$.}
    \begin{tabular}{c|l|rrr}
            \multicolumn{1}{c}{} & Graph & $n$ & $m$ & $\Delta$ \\
            \midrule
            \multirow{17}{*}{\rotatebox[origin = c]{90}{Low-degree graphs}}
            & \textsf{packing} & \numprint{2145839} & \numprint{34976486} & \numprint{18} \\
            & \textsf{channel} & \numprint{4802000} & \numprint{85362744} & \numprint{18} \\
            & \textsf{hugebubbles} & \numprint{19458087} & \numprint{58359528} & \numprint{3} \\
            & \textsf{nlpkkt240} & \numprint{27993600} & \numprint{746478752} & \numprint{27} \\
            & \textsf{europe.osm} & \numprint{50912018} & \numprint{108109320} & \numprint{13} \\
            & \textsf{kmerU1a} & \numprint{64678340} & \numprint{132787258} & \numprint{35} \\
            & $\textsf{rgg}_{\textrm{3D}}26$ & \numprint{67106449} & \numprint{755904090} & \numprint{34} \\
            & $\textsf{rgg}_{\textrm{2D}}26$ & \numprint{67108858} & \numprint{1149107290} & \numprint{45} \\
            & $\textsf{del}_{\textrm{3D}}26$ & \numprint{67108864} & \numprint{1042545824} & \numprint{40} \\
            & $\textsf{del}_{\textrm{2D}}26$ & \numprint{67108864} & \numprint{402653086} & \numprint{26} \\
            & $\textsf{rgg}_{\textrm{3D}}27$ & \numprint{134214672} & \numprint{1575628350} & \numprint{36} \\
            & $\textsf{rgg}_{\textrm{2D}}27$ & \numprint{134217728} & \numprint{2386714970} & \numprint{46} \\
            & $\textsf{del}_{\textrm{2D}}27$ & \numprint{134217728} & \numprint{606413354} & \numprint{14} \\
            & $\textsf{del}_{\textrm{3D}}27$ & \numprint{134217728} & \numprint{2085147648} & \numprint{40} \\
            & \textsf{kmerP1a} & \numprint{138896082} & \numprint{296930692} & \numprint{40} \\
            & \textsf{kmerA2a} & \numprint{170372459} & \numprint{359883478} & \numprint{40} \\
            & \textsf{kmerV1r} & \numprint{214004392} & \numprint{465409664} & \numprint{8} \\
            \midrule
            \multirow{15}{*}{\rotatebox[origin = c]{90}{High-degree graphs}}
            & \textsf{amazon} & \numprint{400727} & \numprint{4699738} & \numprint{2747} \\
            & \textsf{eu-2005} & \numprint{862664} & \numprint{32276936} & \numprint{68963} \\
            & \textsf{youtube} & \numprint{1134890} & \numprint{5975246} & \numprint{28754} \\
            & \textsf{in-2004} & \numprint{1382867} & \numprint{27182946} & \numprint{21869} \\
            & \textsf{com-orkut} & \numprint{3072441} & \numprint{234370166} & \numprint{33313} \\
            & \textsf{enwiki-2013} & \numprint{4203323} & \numprint{183879456} & \numprint{432260} \\
            & \textsf{enwiki-2018} & \numprint{5608705} & \numprint{234488590} & \numprint{248444} \\
            & \textsf{uk-2002} & \numprint{18483186} & \numprint{523574516} & \numprint{194955} \\
            & \textsf{arabic-2005} & \numprint{22743881} & \numprint{1107806146} & \numprint{575628} \\
            & \textsf{uk-2005} & \numprint{39454463} & \numprint{1566054250} & \numprint{1776858} \\
            & \textsf{it-2004} & \numprint{41290648} & \numprint{2054949894} & \numprint{1326744} \\
            & \textsf{twitter-2010} & \numprint{41652230} & \numprint{2405026092} & \numprint{2997487} \\
            & \textsf{sk-2005} & \numprint{50636059} & \numprint{3620126660} & \numprint{8563816} \\
            & \textsf{uk-2007-05} & \numprint{105153952} & \numprint{6603753128} & \numprint{975419} \\
            & \textsf{webbase-2001} & \numprint{115554441} & \numprint{1709619522} & \numprint{816127} \\
        \bottomrule
    \end{tabular}
    \label{tab:graphs}
\end{table}

\fi

\end{document}